\documentstyle[aps]{revtex}
\begin{document} 
\title{\Large Analytical solutions of a Dirac bound state equation
\\ 
and their field interpretation} 
\author{L.Micu\thanks{E-mail address: 
lmicu@theory.nipne.ro}\\Horia Hulubei Institute for Physics and Nuclear
Engineering\\
 Bucharest POB MG-6, 76900 Romania\\
~~~\\Institute for Theoretical Physics of 
the University of Bern\\
CH-3012 Bern, Sidlerstrasse 5}
\maketitle
\begin{abstract}
We solve the single particle Dirac equation with a 
particular confining potential and comment its significance from the
point of view of the quantum field theory.  
We show that the solutions  
describe a complex physical system made of independent constituents: a free
particle and aof an effective field representing the confining potential.
\end{abstract}

\section{The Dirac equation}

In the description of low energy hadronic processes the
so called "QCD inspired" models based on phenomenological notions of
constituent quarks, confining potentials and hyperfine interaction, remain
very useful tools. To understand their connection with QCD 
a first requirement is to express in field language the information
aquired from these models. 
In a recent paper \cite{micu} we have presented a way to do this.
We have shown that in the momentum space
representation a bound system of particles 
can  be treated like a gas of free particles and a
collective excitation of a background field which must be seen as the
stationary time averaged result of a continuous series of 
quantum fluctuations. 

The main features of the method are now discussed on a simple example
which has analytical solutions.
 
We consider the case of a single particle with spin 1/2 confined to a
certain region of space by an external field represented by a particular
scalar-vector combination of linear rising and Coulomb-like potentials
frequently used in quark models \cite{lsg}. The bound state function, 
$\psi$, is an
eigenfunction of the single particle Dirac Hamiltonian and satisfies the 
equation
\begin{equation}\label{h1}
\left(\vec{\alpha}\cdot\vec{p}+\beta m+
{\mathcal
V}(\vec{r})\right)\psi(\vec{r})=\varepsilon~
\psi(\vec{r})
\end{equation} 
where 
\begin{equation}\label{v}
{\mathcal
V}(\vec{r})={\mathcal
V}_\pm(\vec{r})={1\over2}\beta\left({\mathcal
V}_1(\vec{r})+
{\mathcal V}_2(\vec{r})\right)\pm{1\over2}\left({\mathcal
V}_1(\vec{r})-{\mathcal V}_2(\vec{r})\right) 
\end{equation} 
and
\begin{equation}\label{v1}
{\mathcal V}_1(\vec{r})=\zeta~\vert\vec{r}\vert ~~~~{\mathrm
a}{\mathrm
n}{\mathrm d}~~~ {\mathcal V}_2(\vec{r})={\xi\over
\vert\vec{r}\vert}-2m_i+2\sqrt{\zeta\over\xi}~\left(\vec{\sigma}
\cdot\vec{L}+{1\over2}\right). 
\end{equation}

We mention that the potential confinement has been 
studied in connection with the problem of a light quark bound to a
heavy antiquark \cite{ovs} and exact solutions of the single particle
problem have been
found also for scalar-vector oscillator potential \cite{tbw} and for other
combinations of scalar-vector Coulomb-like and linear rising
potentials \cite{jf}. Our solutions are however particularly simple 
and have some nice features which deserve a detailed examination. 

Following 
the usual treatment of the Dirac equation with a central potential \cite{rose}
we write $\psi$ as:
\begin{equation}\label{psi}
\psi(\vec{r})={1\over r}\left(\begin{array}{c}
F(r)~{\mathcal Y}^M_{lJ}\\
iG(r)~{\mathcal Y}^M_{l'J}
\end{array}
\right)
\end{equation}
where $l,~l'$ are either $l=J-{1\over2},~l'=J+{1\over2}$, or 
$l=J+{1\over2},~l'=J-{1\over2}$ and ${\mathcal
Y}^M_{lJ}=\sum_{m,s}~C^{l~~{1\over2}~~J}_{m~s~~M}~Y^m_l({\vec{r}\over 
r})~\chi^s$.

Taking now ${\mathcal V}^{(i)}={\mathcal V}_+$ in (\ref{h1})
and recalling the relations
\begin{equation}\label{pr}
\vec{\alpha}\cdot\vec{p}=\alpha_r~\left[p_r+{i\over
r}~(\vec{J}^2-\vec{L}^2+{1\over 4})\right] 
\end{equation}
and 
\begin{equation}
{\vec{\sigma}\cdot\vec{r}\over r}{\mathcal
Y}^M_{(J\pm{1\over2})~J}=-{\mathcal Y}^M_{(J\mp{1\over2})~J}
\end{equation}
where $\alpha_r={\vec{r}\over r}\vec{\alpha}$,
$p_r=-i{1\over
r}{\partial\over \partial r}r$, $\vec{J}$ is the total angular momentum,
$\vec{L}$ is the orbital momentum 
we replace ec.(\ref{h1}) by two
coupled differential equations  
\begin{equation}\label{sf}
-G'(r)+{J(J+1)-l'(l'+1)+{1\over4}\over r}~G(r)~=~(\varepsilon -m-\zeta
r)~F(r)
\end{equation}
\begin{eqnarray}\label{sg}
&&F'(r)-{J(J+1)-l(l+1)+{1\over 4}\over r}~F(r)=\nonumber\\
&&=~\left[\varepsilon-m+{\xi\over
r}+2\sqrt{\zeta\over\xi}\left(J(J+1)-l'(l'+1)-{1\over
4}\right)\right]~G(r).
\end{eqnarray}

Now we consider the case $l=J-{1\over2},~l'=J+{1\over2}$
and observe that a first condition for $\psi$ to be finite at the origin
is $F(r)\sim r^{1+k}, k\ge0$. From (\ref{sf}) it follows $G(r)\sim
r^{2+k}$ and from (\ref{sg}) one gets $k=l$. We also observe from
(\ref{sg}) that $G'(r)$ behaves like $rF(r)$ at infinity, so we take
\begin{equation}\label{f}
F(r)=r^{l+1}{\mathrm e}^{-\alpha r}~\sum_{i=0}a_i~r^i
\end{equation}
\begin{equation}\label{g}
G(r)=r^{l+2}{\mathrm e}^{-\alpha r}~\sum_{i=0}b_i~r^i.
\end{equation}
Introducing the expressions (\ref{f}) and (\ref{g}) in the equations
(\ref{sf}) and (\ref{sg}) we get the following relations connecting the
coefficients $a_i,~b_i$: 
\begin{equation}\label{i}
-(2l+n+3)b_{n}+\alpha b_{n-1}=(\varepsilon-m)a_{n}-\zeta
a_{n-1} 
\end{equation}
\begin{equation}\label{ii}
na_n-\alpha a_{n-1}=\left[\varepsilon-m-2\sqrt{\zeta\over
\xi}\left(l+{3\over 2}\right)\right]b_{n-2}+\xi b_{n-1}.
\end{equation}
We cut the series in (\ref{f}) at $n=N$ by requiring  $a_{N+k}=0$ for any
$k\ge1$ and obtain from (\ref{ii}) the following
restrictions for $b_{N+k}$: 
\begin{equation}\label{E}
0=\left[\varepsilon-m-2\sqrt{\zeta\over \xi}\left(l+{3\over
2}\right)\right]b_{N+k-1}+\xi
b_{N+k}.
\end{equation}
Equation (\ref{i}) gives for any $k\ge 1$: 
\begin{equation}\label{M}
-(2l+N+k+1)b_{N+k}+\alpha b_{N+k-1}=-\delta_{k1}~\zeta a_{N+k-1}.
\end{equation}
Supposing the series in $G(r)$ is cut at $i=M$ where $M\ge N+1$,
it follows from (\ref{M}) that $b_i=0$ for any $N~< i\le M$.
On the other side, the condition $b_N\ne0$ implies
\begin{equation}\label{q}
\varepsilon-m-2\sqrt{\zeta\over\xi}\left(l+{3\over 2}\right)=0
\end{equation}
which is the quantification condition for the energy levels.
Introducing the condition (\ref{q}) in the equations 
(\ref{i}) and (\ref{ii}) we write them as follows 
\begin{equation}\label{in}
-(2l+n+3)b_{n}+\alpha b_{n-1}=2\sqrt{\zeta\over \xi}\left(l+{3\over
2}\right)a_{n}-\zeta a_{n-1}
\end{equation} 
\begin{equation}\label{iin}
(n+1)a_{n+1}-\alpha a_{n}=\xi b_{n}.
\end{equation}

For $n=0,1,2,...,N$ we get from (\ref{in}) and (\ref{iin})
\begin{eqnarray}\label{set}
&&-(2l+3)b_0=2\sqrt{\zeta\over\xi}\left(l+{3\over 2}\right)a_0\nonumber\\
&&a_1-\alpha a_0=\xi b_0\nonumber\\
&&-2(l+2)b_1+\alpha b_0=2\sqrt{\zeta\over\xi}\left(l+{3\over 2}\right)
a_1-\zeta a_0\nonumber\\ &&2a_2-\alpha a_1=\xi b_1\nonumber\\ 
&&-\alpha a_{N}=\xi b_{N}\nonumber\\
&&\alpha b_N=-\zeta a_N.
\end{eqnarray}

From the last two equalities it follows that
\begin{equation}
\alpha=\sqrt{\zeta\xi}; ~~~~~~~~~~\zeta\xi>0.
\end{equation}  
Introducing these relations into the
first equalities of the set (\ref{set}) we have successively: 
$a_1=0,~b_1=0,~a_2=0,...$ which means that the  solution of the eigenvalue
equation (\ref{h1}) reads  
\begin{equation}\label{psif}
\psi_J^M(\vec{r})=\left(\begin{array}{r} r^{J-{1\over2}}{\mathrm 
e}^{-\sqrt{\zeta\xi}r}~{\mathcal Y}^M_{(J-{1\over2}) J}\\
-i\sqrt{\zeta\over\xi}~r^{J+{1\over 2}}{\mathrm
e}^{-\sqrt{\zeta\xi}r}~{\mathcal Y}^M_{(J+{1\over2}) J}
\end{array} \right) 
\end{equation}
and $\varepsilon_J=m+2\sqrt{\zeta\over\xi}(J+1)$.

Proceeding in a similar manner it can be shown that there is no
solution in the case  
$l=J+{1\over 2},~l'=J-{1\over 2}$,
so (\ref{psif}) is the single one and $\varepsilon_J$ is $(2J+1)$ fold
degenerate. 

We also define
$\bar\psi_J^M=(\psi_J^M)^+\gamma^0$ which satisfies the same
equation as $\psi_J^M$ and the charge conjugate solution
$\psi^{cM}_J$ 
\begin{equation}\label{psic}
\psi^{cM}_J(\vec{r})=i\gamma^2\gamma^0 (\bar{\psi}^M_J)^T(\vec{r})= 
\left(\begin{array}{r}
i\sqrt{\zeta\over\xi}~r^{J+{1\over2}}{\mathrm
e}^{-\sqrt{\zeta\xi}r}~{\mathcal Y}^M_{(J+{1\over2}) J}\\
r^{J-{1\over 2}}{\mathrm e}^{-\sqrt{\zeta\xi}r}~
{\mathcal Y}^M_{(J-{1\over2})J}
\end{array} \right) 
\end{equation}
which satisfies the
equation (\ref{h1}) with the potential ${\mathcal V}_-$
and corresponds to the eigenvalue $\varepsilon^c_J 
=-\varepsilon_J=-m-2\sqrt{\zeta\over\xi}(J+1)$.

\section{The field interpretation of the bound state function}

Our purpose now is to find the relativistic meaning of   
the bound state functions $\psi$ and $\psi^c$ in the hope to create a link 
with the quantum field theory. 

To this end we have first to expand $\psi$ and $\psi^c$ in terms of
the free Dirac solutions, the only ones having a real relativistic
character and then to interpret in a Lorentz covariant manner the
contribution of the confining potential to the bound state energy \cite{micu}.

For reasons of simplicity we shall work with the lowest $J$ functions 
\begin{equation}\label{plow}
\psi^{\rho}_{1\over2}(\vec{r})=\left(\begin{array}{r}
{\mathrm e}^{-\sqrt{\zeta\xi}r}\chi^\rho\\
-i\sqrt{\zeta\over\xi}~r{\mathrm
e}^{-\sqrt{\zeta\xi}r}{\mathcal Y}^\rho_{1{1\over2}}
\end{array} \right)
\end{equation}
and
\begin{equation}\label{pclow}
\psi^{c\rho}_{1\over2}(\vec{r})=\left(\begin{array}{r}
i\sqrt{\zeta\over\xi}~r{\mathrm
e}^{-\sqrt{\zeta\xi}r}~{\mathcal Y}^\rho_{1{1\over2}}\\
{\mathrm e}^{-\sqrt{\zeta\xi}r}~\varphi^\rho
\end{array} \right)
\end{equation}   
where $\chi^\rho$ and $\varphi^\rho$ are two component spinors and project
them on the free Dirac spinors.

In the first case the projection $\phi^{(+)}_{\rho s}(\vec{k})$ on a
positive energy free
state has the following expression  
\begin{eqnarray}\label{puk}
&&\phi^{(+)}_{\rho s}(\vec{k})=\int d^3r{\mathrm e}^{-i\vec{k}\vec{r}}
\bar{u}^s(\vec{k})\psi^\rho_{1\over2}(\vec{r})=\nonumber\\
&&={4\pi\sqrt{\zeta\xi}\over(\zeta\xi+\vec{k}^2)^2}{\sqrt{e+m}
\over\sqrt{2m^3}}\left(1-4\sqrt{\zeta\over\xi} 
{e-m\over\zeta\xi+\vec{k}^2}\right)\delta_{\rho s} 
\end{eqnarray}
where $e=\sqrt{\vec{k}^2+m^2}$ 
and, according to the general principles of the quantum mechanics, it
represents the probability amplitude to find a free particle with momentum
$\vec{k}$ and spin $s$ in the bound state $\psi^\rho_{1\over2}$.

Projecting $\psi$ on a negative energy free Dirac state one obtains
\begin{eqnarray}\label{pvk}
&&\phi^{(-)}_{\rho s}(\vec{k})=\int d^3r {\mathrm e}^{i\vec{k}\vec{r}}
\bar{v}^s(\vec{k})~\psi^\rho_{1\over2}(\vec{r})=\nonumber\\
&&={4\pi\sqrt{\zeta\xi}\over(\zeta\xi+\vec{k}^2)^2}{1\over\sqrt{2m(e+m)}}
\left(1+4\sqrt{\zeta\over \xi}{e+m\over\zeta\xi+\vec{k}^2}\right)
\bar{v}^s(\vec{k})\gamma^0u^\rho(\vec{k}) 
\end{eqnarray}
which shows that the existence of a negative energy free particle is
accompanied by the dissolution into the vaccuum of a
particle-antiparticle pair. We recall that the projections of 
negative energy free states on the free states with positive energy are
allways zero and hence the existence of the nonvanishing result
(\ref{pvk})
has to be considered an effect of a classical potential. This demonstrates
that $\psi$ is not a single particle state in the sense of the classical
quantum mechanics and that its physical content is more complex than that. 
This means also that the part missing from the complete
description of the bound state is related to the
confining potential which is now indissoluble connected to the bound
state function.  

In the attempt to identify this part we notice that the
position vector with respect to the center of forces, $\vec{r}$ can be
written as $\vec{R}-\vec{R}_0$, where
$\vec{R},\vec{R}_0$ are the position vectors of the particle and of
the center of forces with respect to the origin of
the coordinate frame and therefore $\psi$ appears to describe
a system made of two components: a free particle having the
momentum $\vec{k}$ and an additional component, denoted $\Phi$ in
the following, carrying the momentum $\vec{Q}$ defined as: 
\begin{equation}\label{vQ}
\vec{Q}=-\vec{k}
\end{equation}
which is the recoil momentum of the center of forces due to the
motion of the free particle. 

Besides, recalling that the bound state function $\psi$ describes a
stationary
system with the energy $\varepsilon$ we conclude that the additional 
component must carry the energy $Q^0$ which represents the potential
energy of the bound system and is defined as follows: 
\begin{equation}\label{Q}
Q^0=\varepsilon\mp\sqrt{\vec{k}_1^2+m_1^2}.
\end{equation}

It is important to mention that by introducing $\Phi$
as an independent component of the bound state and by defining 
its 4-momentum as a linear combination of free 4-momenta 
(see (\ref{vQ}) and (\ref{Q})) the contribution of the binding potential
to the bound state energy aquires a well defined, Lorentz covariant
significance. The major advantage of this fact is that one can write
immediately a Lorentz covariant representation of the bound state without  
worrying about the transformation properties of the binding
potential at boosts. The single change to be made is to write the
functions $\phi^{(\pm)}$ in a Lorentz covariant form which is
obtained by replacing $\vec{k}$ by $k^\mu_T=k^\mu-(\pi\cdot k)\pi^\mu$
where the vector
$\pi^\mu=({1\over\sqrt{1-\omega^2}},{1\over\sqrt{1-\omega^2}}\vec{\omega})$
is used to write the Lorentz transformation from the initial frame 
where the confining potential is defined 
to a frame moving with the velocity $\vec{\omega}$ with respect to it.

We have:
\begin{eqnarray} 
&&\phi^{(+)}_{\rho s}(k)={4\pi\sqrt{\zeta\xi}\over (\zeta\xi+(\pi\cdot
k)^2-m^2)^2} {\sqrt{(\pi\cdot k)-m}\over m\sqrt{2m}} 
\left(1- 4\sqrt{\zeta\over\xi}{(\pi\cdot
k)-m\over\zeta\xi+(\pi\cdot k)^2-m^2}\right)\delta_{\rho s},\nonumber\\
&&\phi^{(-)}_{\rho s}(k)=
{4\pi\sqrt{\zeta\xi}\over (\zeta\xi+(\pi\cdot
k)^2-m^2)^2} ({1\over\sqrt{2m((\pi\cdot k)+m)}} 
\left(1+ 4\sqrt{\zeta\over\xi}{(\pi\cdot
k)-m\over\zeta\xi+(\pi\cdot
k)^2-m^2}\right)\nonumber\\
&&\times\bar{v}^s(k)(\pi\cdot
\gamma)u^\rho(k).
\end{eqnarray} 

Then, in field notation, the stationary Lorentz covariant expression of
the bound state function $\psi$ can be written:
\begin{eqnarray}\label{Psi} 
&&\psi^\rho(\vec{R},\vec{R}_0,t)=\nonumber\\
&&\sum_{\rho}\int d^3k{m\over
e}d^4Q~\left(\delta^{(4)}(Q+k-P)~
\phi^{(+)}_{\rho s} (k)
a_s(k)\Phi(Q)~u^{s}(k){\mathrm
e}^{-i(e+Q^0)t+i\vec{k}\vec{R}+i\vec{Q}\vec{R}_0}\right.\nonumber\\
&&\left.-\delta^{(4)}(Q-k-P)~\phi^{(-)}_{\rho s} (k)
b^\dagger_s(k)\Phi(Q)~v^{s}(k){\mathrm
e}^{i(e-Q^0)t-i\vec{k}\vec{R}+i\vec{Q}\vec{R}_0}\right)
\end{eqnarray}
where the annihilation operator of a negative energy free state
$a_-$ has been replaced by the creation operator of an
antiparticle, $b^\dagger$. The four components of the momentum
$P^\mu$ are $E={\varepsilon\over\sqrt{1-\omega^2}}$ and 
$\vec{P}=-\vec{\omega}E$.
$\Phi$ designates the additional component representing the confining
potential 
and, as it can be seen from (\ref{Psi}), it must be seen as a reservoir of
particles and energy in the sense the vaccuum is for the free Dirac
equation.

Similar results are obtained in the case of the charge conjugate function
where one writes
\begin{eqnarray}\label{cPsi}
&&\psi^{c\rho}(\vec{R},\vec{R}_0,t)=\nonumber\\
&&\sum_{\rho}
\int d^3k{m\over
e}d^4Q~\left(\delta^{(4)}(k+Q-P)
\phi^{c(-)}_{\rho s} (k)
a_s(k)\Phi(Q)~u^{s}(k){\mathrm
e}^{-i(e+Q^0)t+i\vec{k}\vec{R}+i\vec{Q}\vec{R}_0}\right.\nonumber\\
&&\left.-\delta^{(4)}(Q-P-k)\phi^{c(+)}_{\rho s}(k)
b^\dagger_s(k)\Phi(Q)~v^{s}(k){\mathrm
e}^{i(e-Q^0)t-i\vec{k}\vec{R}+i\vec{Q}\vec{R}_0}\right)
\end{eqnarray}
where $\phi^{c(+)}$ and $\phi^{c(-)}$ can be obtained
from $\phi^{(+)}$ and $\phi^{(-)}$ respectively by performing the
replacement $u^r\leftrightarrow-v^r$.

Concluding this paper we notice that from the point of view of the
field theory 
a particle in a bound state can be seen as a system made of two
components: a free particle and an effective field $\Phi$ which
has a double face: in momentum space it is a reservoir
of particles and energy while in coordinate representation it 
represents a kind of a box where the particle is confined.

This image is similar to that of bag models \cite{bm}. It is a
time averaged image, not an instantaneous one and it is expected to hold
whenever the observation time is longer or at least equal to the time 
giving a stable average.

We also notice that, as suggested by the relativistic interpretation of
the bound state functions (\ref{Psi}) and (\ref{cPsi}), the Dirac equation
with a confining potential is a field equation and its solutions represent
a complex physical structure where the free positive and negative 
energy states are mixed.

\vskip0.5cm
Acknowledgments The author thanks Fl. Stancu for valuable
suggestions and continuous encouragement.

The work has been completed during author's visit at ITP
of the University of Bern under the SCOPE Programme.   
The warm hospitality at ITP and the financial support from the Swiss
National Science Foundation are gratefully acknowledged.

\end{document}